\documentclass[11pt]{amsart}
\usepackage{amssymb}
\usepackage{amscd}
\numberwithin{equation}{section}
\newtheorem{claim}{}[section]

\newtheorem{theorem}[claim]{Theorem}

\newtheorem{corollary}[claim]{Corollary}

\begin{document}

\dedicatory{Dedicated to Gert Pedersen, who is missed for both his brilliance \\and his exuberant sense of humor.}

\title[Matrix Convexity]{A Matrix Convexity Approach\\to Some Celebrated Quantum Inequalities}
\author{Edward G. Effros}
\thanks{Supported by the National Science Foundation DMS-0100883}
\address{Department of Mathematics\\UCLA, Los Angeles, CA 90095-1555}
\email[Edward G. Effros]{ege@math.ucla.edu}

\date{January 28, 2008}
\maketitle
\begin{abstract} Some of the important inequalities associated with quantum entropy are immediate algebraic consequences of the Hansen-Peder\-sen-Jensen inequalities. A general argument is given using matrix perspectives of operator convex functions. A matrix analogue of Mar\'{e}chal's extended perspectives provides additional inequalities, including a $p+q\leq 1$ result of Lieb.
\end{abstract}

\section{Introduction}

In 1973, Elliott Lieb published a ground-breaking paper on operator inequalities \cite{Lie}. This and a subsequent paper by  Lieb and Ruskai \cite{LiR} have had a profound effect on quantum statistical mechanics, and more recently on quantum information theory. Since then, a number of attempts have been made to elucidate and extend these results. Two particularly elegant examples are those of Nielsen and Petz \cite{Nie}, and Ruskai \cite{Rus}), which use the analytic representations for operator convex functions. On the other hand, Frank Hansen \cite {Ha2} has developed a powerful theory that utilizes {\em geometric means} of positive operators. The latter noton was formulated by Pusz and Woronowicz \cite{Wo}, and subsequently investigated by Ando \cite{Ando} (see the discussion in Section 3) and by  Kubo and Ando \cite{Ka}. 

Here we present what is arguably the simplest approach to these inequalities. This is accomplished by using matrix analogues of two elementary ideas from classical convexity theory: the Jensen inequality, and the construction of the perspective of a convex function. For the first, we employ the matricial Jensen inequality of Frank Hansen and Gert Pedersen \cite{Ha1}, \cite{Ha}. As we point out in Section 5, the affine and  homogeneous versions of this inequality can be proved in a relatively few lines drawn from those papers. The non-commutative analogues of perspectives are completely straight-forward in the context of the left and right module operations that are standard to the subject. In section 4 we show that the same approach may be used to quantize Mar\'{e}chal's
extended version of the perspective. We apply this to prove Lieb's generalized $p+q\leq 1$ inequality (see also the elegant proof in \cite{Ha2}).

The appearance of notions from convexity theory suggests that other geometric techniques will prove useful in the operator context. In a different direction, quantum information theory is likely to have an impact on the theory of matrix convexity. This possility is considered in Section 4.

I am grateful to Frank Hansen for alerting us to his work in this area. I also wish to thank Mary Beth Ruskai, who corrected a number of errors in my first manuscript, Jon Tyson for a host of suggestions, and Richard Kadison for his encouragement.

Since the basic difficulties are already apparent in finite
dimensions, we have restricted our attention to finite matrices, and we have avoided any attempt at full
generality even in that context.

\section{The classical and matrix notions of perspectives}

Given a convex function $f$ defined on a convex set $K\subseteq \Bbb{R}^{n}$, the \emph{perspective} $g$ is defined on the subset 
\begin{equation*}
L=\left\{ (x,t):t>0\text{ and }x/t\in K\right\} 
\end{equation*}
by 
\begin{equation*}
g(x,t)=f(x/t)t
\end{equation*} 
(see \cite{Hu}). It is a simple exercise to verify that $g(x,t)$ is a jointly  convex function in the sense that if $0\leq c \leq 1$, then
\[
g(cx_{1}+(1-c)x_{2},ct_{1}+(1-c)t_{2})
\leq cg(x_{1},t_{1})+(1-c)g(x_{2},t_{2}). 
\]
An elementary but important example is provided by the continuous convex function
$f(x)=x\log x,$ with $f(0)=0$ defined on $[0,\infty)\subseteq \Bbb{R}$. It follows that the perspective function 
\[
g(x,t)=t\frac{x}{t}\log \frac{x}{t}=x\log x-x\log t
\]
is jointly convex. Letting $p=(p_{i})$ and $q=(q_{i})$ be finite
probability measures with $p_{i}>0$ and $q_{i}>0,$ the convexity of $f$ implies that the classical entropy
\begin{equation*}
H(p)=-\sum p_{i}\log p_{i}
\end{equation*}
is concave, and the convexity of $g$ implies that the relative entropy
\begin{equation*}
(q,p)\mapsto H(q||p)=\sum p_{i}\log p_{i}-p_{i}\log q_{i}
\end{equation*}
is jointly convex on pairs of probability measures.

We recall that if $f:I=[a,b]\rightarrow \Bbb{R}$ is continuous, and $T$ is an $n\times n$ self-adjoint matrix with spectrum in $[a,b]$, then we can define $%
f_{n}(T)$ by spectral theory (or by using a basis in which $T$ is diagonal). $f$ is said
to be \emph{matrix convex} if for each $n\in \Bbb{N},$ the corresponding
function $f_{n}$ is convex on the self-adjoint $n\times n$ matrices with
spectrum in $[a,b]$. Throughout the rest of the paper we only consider $n\times n$ matrices, and we usually omit the subscript $n$. The following is the affine version of the Hansen-Pedersen-Jensen inequality \cite{Ha} (see Section 5).  

\begin{theorem}If $f$ is matrix convex, and $A$ and $B$ satisfy $A^{*}A+B^{*}B=I_{n},$ then 
\begin{equation}
f(A^{*}T_{1}A+B^{*}T_{2}B)\leq A^{*}f(T_{1})A+B^{*}f(T_{2})B.
\end{equation}
\end{theorem}

We begin with some matrix conventions. Given matrices $L$ and $R$,
we let $[L,R]=LR-RL$. Let us suppose that $L>0$ and $R>0$. If $[L,R]=0$, i.e., the matrices commute, then we may find a basis in which both matrices are diagonalized. It follows that $LR>0$, $[L,R^{-1}]=0$, and we may unambiguously write $\frac{L}{R}$ for the quotient. We also recall that for any continuous function $f,$ $%
f(L)$ commutes with any operator commuting with $L$ (including $L$ itself). Using simultaneously diagonalized matrices, it is evident that we have relations such as $\log LR^{-1} = \log L - \log R$.

\begin{theorem} Suppose that $f$ is operator convex. When restricted to positve commuting matrices $L,R$, the ``perspective function'' 
\begin{equation}
(L,R)\mapsto g(L,R)=f\left( \frac{L}{R}\right) R
\end{equation}
is jointly convex in the sense that if  $L=cL_{1}+(1-c)L_{2}$ and $R=cR_{1}+(1-c)R_{2}$ where $[L_{j},R_{j}]=0$ ($j=1,2$), and $0 \leq c \leq 1$, then 
\begin{equation}
g(L,R)\leq cg(L_{1},R_{1})+(1-c)g(L_{2},R_{2}).
\end{equation}
\end{theorem}

\proof The matrices $A=(cR_{1})^{1/2}R^{-1/2}$ and $%
B=((1-c)R_{2})^{1/2}R^{-1/2}$ satisfy $A^{*}A+B^{*}B=I.$ From Theorem 2.1, 
\begin{eqnarray*}
\lefteqn{g(L,R)}\\
&=& Rf\left(\frac{L}{R}\right)\\
&=&R^{1/2}f(R^{-1/2}LR^{-1/2})R^{1/2} \\
&=&R^{1/2}f\left( A^{*}\left( \frac{L_{1}}{R_{1}}\right) A+B^{*}\left( \frac{%
L_{2}}{R_{2}}\right) B\right) R^{1/2} \\
&\leq &R^{1/2}\left( A^{*}f\left( \frac{L_{1}}{R_{1}}\right) A+B^{*}f\left( 
\frac{L_{2}}{R_{2}}\right) B\right) R^{1/2} \\
&=&(cR_{1})^{1/2}f\left( \frac{L_{1}}{R_{1}}\right)
(cR_{1})^{1/2}+((1-c)R_{2})^{1/2}f\left( \frac{L_{2}}{R_{2}}\right)
((1-c)R_{2})^{1/2} \\
&=&cg(L_{1},R_{1})+(1-c)g(L_{2},R_{2}).
\end{eqnarray*}
\endproof

The following result is due to Lieb and Ruskai \cite{LiR} (a related early discussion may be found in Lindblad \cite{Lin}).

\begin{corollary}The relative entropy function
\begin{equation*}
(\rho,\sigma )\mapsto S(\rho ||\sigma )=\mathrm{Trace}\,\rho \log \rho -\rho \log \sigma 
\end{equation*}
is jointly convex on the strictly positive $n\times n$ density matrices $%
\rho ,\sigma $.
\end{corollary}

\proof We let $M_{n}$ have the usual Hilbert space structure determined by $%
\langle X,Y\rangle =\mathrm{Trace}$ $XY^{*}.$ Given positive density
matrices $\sigma $ and $\rho ,$ we define operators $R$ and $L$ on $M_{n}$
by $L(X)=\rho X$ and $R(X)=X\sigma .$ Then we have that $L(X)$ and $R(X)$ are
commuting positive operators on the Hilbert space $M_{n}.$ On the other hand
the function $f(x)=x\log x$ is operator convex (see \cite{Bha}, p. 123), and thus
\begin{eqnarray*}
\langle g(L,R)(I),I\rangle &=&\left\langle R\left(\frac{L}{R}\right)\log \left(\frac{L}{R}%
\right)(I),I\right\rangle \\&=&\langle L(\log L-\log R)(I),I\rangle \\
 &=&\mathrm{Trace} \rho\log \rho -\rho \log \sigma =S(\rho ||\sigma )
\end{eqnarray*}
is jointly convex. 
\endproof
The following is due to Lieb \cite{Lie}. It was subsequently used by Lieb and Ruskai to prove strong subadditivity for relative entropy \cite{LiR}. A stronger result of Lieb is discussed in the next section.

\begin{corollary} If $0<s<1,$ then the function
\begin{equation*}
F(A,B)=\mathrm{Trace}\,A^{s}K^{*}B^{1-s}K
\end{equation*}
is jointly concave on the strictly positive $n\times n$ matrices $A,B$.
\end{corollary}
\proof Since $f(t)=-t^{s}$ is operator convex (see \cite{Bha} Th.5.1.9),  $g(L,R)=-L^{s}R^{1-s}$ is jointly convex for appropriately commuting
operators. Again using the Hilbert space structure on $M_{n},$ we let $L(X)=AX$ and $R(X)=XB.$ It follows that 

\begin{equation*}
(A,B)\mapsto -\mathrm{Trace}\,A^{s}K^{*}B^{1-s}K=\langle g(L,R)(K^{*}),K^{*}\rangle 
\end{equation*}
is jointly convex.
\endproof

Various generalized entropies may be handled in much the same manner.

\section{Mar\'{e}chal's perspectives}
	
P. Mar\'{e}chal has recently introduced an interesting generalization of perspectivity for convex functions \cite{Ma1}, \cite{Ma2}. This also has a natural matrix version. For this purpose we use the subhomogeneous form of the Hansen-Pedersen-Jensen inequality \cite{Ha1} (see Section 5). We assume that the functions $f$ and $g$ are defined on an interval $I\subseteq {\Bbb R}$, and that $0 \in I$.

\begin{theorem}  If $f$ is matrix convex, and $f(0)\leq 0,$ and that $A$ and $B$
are matrices with $A^{*}A+B^{*}B\leq I_{n},$ then 
\begin{equation*}
f(A^{*}T_{1}A+B^{*}T_{2}B)\leq A^{*}f(T_{1})A + B^{*}f(T_{2})B.
\end{equation*}
\end{theorem}

Given continuous functions $f$ and $h$, and commuting positive matrices $L$ and $R,$
we define
\begin{equation*}
(f\Delta h)(L,R)=f\left(\frac{L}{h(R)}\right)h(R)
\end{equation*}

A close variation of the following result was proved for operator monotone functions $f$ on $(0,\infty)$ by Ando (see \cite{Ando} Theorem 6). His construction (without the extra function $\Phi_{2}$ which can be incorporated with a compostion) is related to Mar\'{e}chal's operation $f\nabla h$ for concave functions $f$ and $h$. Ando invoked the integral representation for operator monotone functions, rather than the matrix convexity argument used below.

\begin{theorem} Suppose that $f$ is matrix convex, $f(0)\leq 0$ and that $h$ is matrix concave with $h>0.$ Then $(L,R)\mapsto(f\Delta h)(L,R)$ is jointly
convex on postive commuting matrices $L,R$ in the sense of Theorem 2.2. 
\end{theorem}
\proof Let us suppose that $L=cL_{1}+(1-c)L_{2}$ and $R=cR_{1}+(1-c)R_{2}$
where $[L_{j},R_{j}]=0$. Then $ch(R_{1})+(1-c)h(R_{2})\leq %
h(R),$ hence 
\begin{eqnarray*}
A &=&c^{1/2}h(R_{1})^{1/2}h(R)^{-1/2} \\
B &=&(1-c)^{1/2}h(R_{2})^{1/2}h(R)^{-1/2}
\end{eqnarray*}
satisfy 
\begin{eqnarray*}
\lefteqn{A^{*}A+B^{*}B}\\
&=&h(R)^{-1/2}ch(R_{1})h(R)^{1/2}+h(R)^{-1/2}(1-c)h(R_{2})h(R)^{-1/2}\\
&\leq& h(R)^{-1/2}h(R)h(R)^{-1/2}I=I.
\end{eqnarray*}
It follows from Theorem 3.1 that  
\begin{eqnarray*}
\lefteqn{(f\Delta h)(L,R)}\\
&=&h(R)^{1/2}f(h(R)^{-1/2}Lh(R)^{-1/2})h(R)^{1/2} \\
&=&h(R)^{1/2}f\left(A^{*}\left( \frac{L_{1}}{h(R_{1})}\right)
A+B^{*}\left( \frac{L_{2}}{h(R_{2})}\right) B\right) h(R)^{1/2} \\
&\leq & h(R)^{1/2}A^{*}f\left( \frac{L_{1}}{h(R_{1})}\right)
Ah(R)^{1/2}+h(R)^{1/2}B^{*}f\left( \frac{L_{2}}{h(R_{2})}\right) Bh(R)^{1/2}
\\
&=&ch(R_{1})^{1/2}f\left( \frac{L_{1}}{h(R_{1})}\right)
h(R_{1})^{1/2}+(1-c)h(R_{2})^{1/2}f\left( \frac{L_{2}}{h(R_{2})}\right)
h(R_{2})^{1/2} \\
&=&c(f\Delta h)(L_{1},R_{1})+(1-c)(f\Delta h)(L_{2},R_{2}).
\end{eqnarray*}
\endproof
To illustrate this result, we reprove Lieb's extension of Corollary 2.4 \cite{Lie}.  

\begin{corollary} Suppose that $0<p,q$ and that $p+q\leq 1$. Then the function
\begin{equation*}
(A,B)\mapsto \mathrm{Trace}\,A^{q}X^{*}B^{p}X
\end{equation*}
is jointly concave on the positive $n\times n$ matrices.
\end{corollary}

\proof Since $p+q\leq 1$, $p+q$ is a convex combination of $q$ and $1$, i.e., we may choose $0\leq t\leq1$ with $p+q=(1-t)q+t1$. If we let $q=s$, then 
\[
p=-tq+t=(1-q)t=(1-s)t.
\] 
 Thus it suffices to show that if $0\leq s,t \leq 1$, then
\begin{equation*}
(A,B)\mapsto -\mathrm{Trace}\,A^{s}X^{*}B^{(1-s)t}X
\end{equation*}
is jointly convex. The functions $f(x)=-x^{s}$ and $h(y)=y^{t}$ are operator convex and concave, respectively, and 
\[
(f\Delta h)(L,R)=h(R)f\left(\frac{L}{h(R)}\right)=-R^{t}\frac{L^{s}}{R^{st}}=-L^{s}R^{(1-s)t}.
\]
 If we let 
 $L(X)=AX$ and $R(X)=XB$ for $X\in M_{n},$ then from Theorem 3.2,\begin{equation*}
(A,B)\mapsto -\mathrm{Trace}\,A^{s}X^{*}B^{(1-s)t}X=\langle (f\Delta h)(L,R)(X^{*}),X^{*}%
\rangle
\end{equation*}
is jointly convex.\endproof

\section{matrix convexity}

Perhaps the most intriguing aspect of Mar\'{e}chal's construction is that it behaves well under the Fenchel-Legendre transform, and under iteration. S{\o}ren Winkler formulated an analogue of the Fenchel-Legendre duality for matrix convex functions \cite{Wi}, but the transforms are generally set-valued mappings. Further progress might result if one could reformulate his theory in terms of "left-right" commuting pairs. It should also be noted that other constructions in classical convexity theory, such as the linear fractional transformations of convex functions (see \cite{B}) might also have matrix generalizations.

Until recently the theory of matrix convexity has suffered from a lack of examples and applications. With the advent of quantum information theory (QIT), this situation has dramatically changed. QIT provides a wealth of remarkable, purely non-classical techniques that might clarify some of the conceptual problems in matrix convexity theory. On the other hand, it seems likely that matrix convexity and more generally non-commutative functional analysis will provide an appropriate framework for many of the calculations in QIT. A striking illustration of this phenomenon can be found in \cite{Dev}.

\section{A brief guide to the \\Hansen-Pedersen-Jensen Inequalities}

The original proof of Theorem 3.1 may be found in \cite{Ha1} (Theorem 2.1). It is both elegant and concise. For our purposes we only need (i) implies (iii) in their proof. On the other hand, Winkler pointed out in \cite{Wi} that Theorem 2.1 is easily derived from Theorem 3.1.
Since our situation is slightly different, we include the argument. 

We fix a point $c\in I$ and define $F(t)=f(t+c)-f(c)$. Given $T=T^{*}\in M_{n}$, we may choose a basis with respect to which $T=\rm{diag}(\lambda_{1},\ldots,\lambda_{n})$. Then 

\begin{eqnarray*}
F(T)&=&F\left( \left[ 
\begin{array}{lll}
\lambda _{1} &  &  \\ 
 & \ddots  &  \\ 
&  & \lambda _{n}
\end{array}
\right] \right)\\ 
&=& \left[ 
\begin{array}{lll}
f(\lambda _{1}+c)-f(c) &  &  \\ 
& \ddots  &  \\ 
 &  & f(\lambda _{n}+c)-f(c)
\end{array}
\right] \\ &=&f(T+cI)-f(c)I
\end{eqnarray*}
is matrix convex and $F(0)=0$. From Theorem 2.1, 
\[
F(A^{*}T_{1}A+B^{*}T_{2}B)\leq A^{*}F(T_{1})A + B^{*}F(T_{2})B,
\] 
\begin{eqnarray*}\lefteqn{f(A^{*}T_{1}A+B^{*}T_{2}B)-f(c)I}\\
&\leq & 
A^{*}f(T_{1})A-f(c)A^{*}A+B^{*}f(T_{2})B-f(c)B^{*}B,
\end{eqnarray*}
and thus
\[
f(A^{*}T_{1}A+B^{*}T_{2}B)\leq A^{*}f(T_{1})A+B^{*}f(T_{2})B.
\]

As pointed out by Winkler \cite{Wi}, the result may be extended to rectangular matrices $A$ and $B$. He used the case $B=0$ to show that a real function $f$ on an interval in $\Bbb{R}$ is a matrix convex function if and only if the supergraphs of the $f_{n}$ form a matrix convex system of sets.

\end{document}